\documentclass[preprint]{article}

\PassOptionsToPackage{numbers,sort&compress}{natbib}

\usepackage{neurips_2025}

\usepackage{rotating}
\usepackage[backref]{hyperref} 
\hypersetup{
    colorlinks,
    linkcolor={purple},
    citecolor={blue!50!black}
}
\usepackage[utf8]{inputenc}    
\usepackage[T1]{fontenc}       
\usepackage{graphicx}
\usepackage{url}               
\usepackage{booktabs}          
\usepackage{colortbl}          
\usepackage{multirow}          
\usepackage{tablefootnote}     
\usepackage{amsfonts}          
\usepackage{amsmath}           
\usepackage{mathtools}         
\usepackage{nicefrac}          
\usepackage{microtype}         
\usepackage{xcolor}            
\usepackage{xspace}            
\usepackage{wrapfig}
\usepackage{enumitem}
\usepackage{adjustbox}      
\usepackage{doi}               
\usepackage{titletoc}          
\usepackage[labelfont=bf]{caption}
\usepackage[capitalise]{cleveref}          
\usepackage{tabularray}
\usepackage{graphicx}
\usepackage[all]{nowidow}
\usepackage{float}
\usepackage{url}
\usepackage{algorithm,algpseudocode}
\usepackage{xcolor}
\usepackage{amsmath}

\usepackage{amsthm} 
\definecolor{myred}{rgb}{1, 0.051, 0.341}
\definecolor{myblue}{rgb}{0.118, 0.533, 0.898}
\definecolor{color_03}{rgb}{0.9843, 0.7372, 0.0156}
\definecolor{color_05}{rgb}{0.890, 0.45490, 0}
\definecolor{color_07}{rgb}{0.9176, 0.2627, 0.2078}

\title{X-ray transferable polyrepresentation learning}

\author{%
  Weronika Hryniewska-Guzik \\
  Warsaw University of Technology \\
  \texttt{weronika.hryniewska.dokt@pw.edu.pl} \\
  \And
  Przemyslaw Biecek \\
  University of Warsaw \\
  Warsaw University of Technology \\
}

\begin{document}

\maketitle
%
%

\begin{abstract}
The success of machine learning algorithms is inherently related to the extraction of meaningful features, as they play a pivotal role in the performance of these algorithms. Central to this challenge is the~quality of data representation. However, the ability to generalize and extract these features effectively from unseen datasets is also crucial.
In light of this, we introduce a novel concept: the polyrepresentation. Polyrepresentation integrates multiple representations of the~same modality extracted from distinct sources, for example, vector embeddings from the Siamese Network, self-supervised models, and interpretable radiomic features. This approach yields better performance metrics compared to relying on a single representation. Additionally, in the context of X-ray images, we demonstrate the~transferability of the created polyrepresentation to a~smaller dataset, underscoring its potential as a pragmatic and resource-efficient approach in various image-related solutions. It is worth noting that the concept of polyprepresentation on the example of medical data can also be applied to other domains, showcasing its versatility and broad potential impact.
\end{abstract}

\section*{Introduction}

The performance of machine learning models depends significantly on the~selection of data representations on which they are applied \cite{repr}. Moreover, also in~artificial intelligence (AI) applied in~radiology, an~increasing number of models, and consequently representations, are emerging \cite{Zhou2022, checklist, Zhang2021}. Unfortunately, most of them are not easily transferable and thus difficult to use \cite{Weiss2016}. In~addition, while attempting transfer learning, the~difficulty in~assessing the~relatedness between a~particular source and target domain can be amplified by the~different ways in~which data is labeled. Consequently, despite the~potential of radiological images, it is not fully understood how the~resulting representations combine and complement each other and how to work with them.


Besides, there is still a~lack of a~multi-faceted approach to solving classification problems more globally. Usually, the~image classification task solution focuses on training a~single deep learning model \cite{duplic, Xie2018, Matsuyama2020}. A~single model solving a~specific task shows only one perspective and does not capture the~whole nature of the~data. 

Model ensembling has great advantages in overcoming many problems of single-model classification \cite{ensembleadversarial}. However, it often looks at the~same data with models of similar low-level feature extractors. Hence, finding a~completely different approach to looking at the~data would be beneficial. It seems useful to statistically describe the~results returned by another model. For example, such a~solution is PyRadiomics library \cite{pyradiomics}, which helps extract features from segmentation masks.

In the~field of natural language processing (NLP), word, sentence, and document embeddings have proven to be a~breakthrough representation of semantic meaning \cite{wordembedding}. In the realm of computer vision, embeddings have emerged as a pivotal concept, particularly through techniques like Siamese Networks. These networks excel at creating vector representations that encapsulate image content effectively, offering a compact yet rich encoding that finds utility in various tasks. One remarkable recent advancement in this arena is DINO \cite{oquab2023dinov2}, which leverages embeddings for self-supervised learning. DINO extends the capabilities of embeddings by fostering the development of discriminative and semantically meaningful representations through data augmentation and negative sampling strategies. This approach showcases the transformative potential of embeddings, as generated by Siamese Networks, in shaping the future of computer vision methodologies, enabling improved generalization and performance across a wide spectrum of applications.

So far, most research studies consider either representation obtained from training neural networks on single-modality data (single models, ensemble models, multi-task models) or representation of multimodal data by training machine learning models on combined data \cite{multimodal}, for example, image embeddings and tabular data. For data of the~same modality, there is still a~scarcity of solutions combining representations extracted from neural networks with representations that rely on expert domain knowledge or engineered hard-coded features. This type of representation we will call polyrepresentation.

\begin{figure*}[t]
\includegraphics[width=\linewidth]{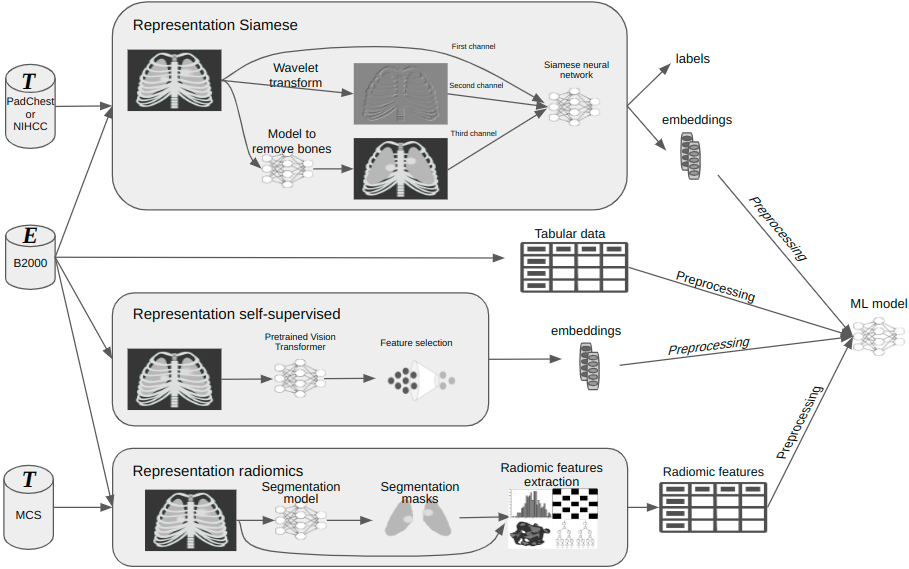}

\caption{An overview of our polyrepresentation creation process, a~combination of Siamese, Self-supervised, and Radiomics representation. Databases marked with $\boldsymbol{T}$ were used for training and marked with $\boldsymbol{E}$ for evaluation. Detailed schema of Siamese Network is in Fig. \ref{figure:siamese}, and of feature importance in Fig. \ref{fig:aspect}.}
\label{fig:schema}
\end{figure*}

In this paper, we introduce a novel concept: the polyrepresentation. This polyrepresentation, depicted in Fig. \ref{fig:schema}, seamlessly integrates multiple representations derived from distinct sources: vector embeddings extracted by the Siamese Network, self-supervised networks, and interpretable radiomic features obtained from the original images and masks generated by a trained segmentation model. What makes this approach particularly noteworthy is its modular nature; individual representation modules can be activated or deactivated at will. A remarkable advantage is that adapting the polyrepresentation entails retraining a classic machine learning model rather than a deep learning model, substantially expediting the process and minimizing computational requirements. Notably, we demonstrate the~possibility of transferring created polyrepresentation to a~smaller dataset, underscoring its potential as a pragmatic and resource-efficient solution in various image-related solutions.

Furthermore, we propose the~3-channel initialization of medical images, which makes the~Siamese Network learn better. In addition to the original image, our solution takes the input image channels preprocessed by the algorithm and neural network.

Our contributions are the~following: (1) we demonstrate that a~multimodal solution that combines three different representations of image data with tabular data results in better results than a~solution based on a~single modality; (2) we show that polyrepresentation - combined representations trained on different tasks lead to higher performance than trained single model; (3) we use the~potential of three channels for grayscale images by adding more information during training that leads to higher performance. We apply the~bones removal model as a~new preprocessing method that initialized one channel of an~image; (4) we check the~contribution of individual channels, which show higher results for channels initialized with a~boneless image and a~wavelet-transformed image.

\section*{Related works}

\hspace{\parindent} \textbf{Siamese Network} \\
The potential of Siamese Networks were recognized in~many papers \cite{Shorfuzzaman2021, yang2020, SiameseXML}. They are particularly useful in~papers related to zero-shot learning, whose aim is to differentiate the~unseen class from classes in~the training dataset. Shorfuzzaman et al. \cite{Shorfuzzaman2021} used Siamese Networks to diagnose diseases in~lung X-ray images. They adapted to this task twin neural network with contrastive loss and trained solution on a~multi-class dataset. In a prior study \cite{yang2020}, a Siamese Network was effectively employed to tackle multi-label data challenges. In~addition, this neural network worked as a multi-task solution, also performing the~classification task. However, it is worth noting that none of these papers use triplets of images; instead, they focus on pairs. As far as we know, the algorithm working with multi-label data is still in the research area.

\textbf{Medical grayscale images} \\
When working with medical images, it is crucial not to introduce bias during the~preprocessing or augmentation \cite{checklist}. For grayscale images, as a~preprocessing method, many works copied one channel of the~image to the~other two channels \cite{duplic}. Nevertheless, for monochromatic medical images, it was shown that using models pre-trained on Imagenet with 3-channel images led to worse results than training the~model on 1-channel grayscale images \cite{Xie2018}. Matsuyama \cite{Matsuyama2020} showed the~possibility of using the~potential of three channels for monochromatic images to obtain a~better model's performance than on 1-channel images. They initialized 3-channel images with preprocessing results for the~vertical, horizontal, and diagonal wavelet coefficients. We also used the~wavelet-transformed data but averaged the~results from the~different axes and put them in~one image's channel.

\textbf{Radiomics features} \\
Recently, radiomics, which is the~extraction of quantitative features from radiological images such as X-ray or CT, has been examined in a~broad range of clinical applications \cite{Horng2022}. Papers have demonstrated the~use of radiomic features for prostate MRI~\cite{Schwier2019}, breast cancer MRI \cite{Granzier2020}, lung CT \cite{Horng2022}, and resectable pancreatic ductal adenocarcinoma CT \cite{Zhang2021}. Zhang et al. \cite{Zhang2021} discusses the added value of two representations for CT images - derived from a~CNN model and radiomic features. He notes that they can be complementary and that combining representations can lead to better models.  We generalize their approach, introduce the term polyrepresentation, add another representation, and tabular data. Moreover, to analyze the~dwarfism condition, a~representation using radiomics features and tabular data was proposed \cite{Qiu2022}. They used two multimodal sources of data. However, since they used only one image representation - radiomic features, it cannot be called a~polyrepresentation. In all of these papers, neither a~transferability of representation was checked nor a~deep learning segmentation model was trained.

\textbf{Transferable representation} \\
Getting a~meaningful image representation is an~important issue~\cite{repr}. Frequently used transfer learning involves deriving representations from a~neural network trained on a~more extensive dataset to extract meaningful features and use them when training on a~smaller dataset \cite{Morano2020}. Thus, a~good representation is demonstrated by adapting to diverse and unseen tasks with the~availability of only a~few examples~\cite{vtab, oquab2023dinov2}. Another approach that produces a~good representation is multi-tasking \cite{He2020}. Additionally, combining textual and visual representations has yielded excellent results in many multimodal tasks, such as visual question answering and image captioning \cite{LopezDeLacalle2020}. In order to create the~best possible representation, we decided to use the~aforementioned approaches. In creating one representation, we used multi-tasking and transfer learning while training our Siamese Network, and in creating the other, we used the features extracted by the~model capable of creating universal data representations DINO.

\section*{X-ray representation based on Siamese Network} \label{sec:siamese}

\subsection*{Datasets selection}

To compare transferability between datasets, we selected two large multi-label datasets: PadChest \cite{bustos2019padchest} and NIHCC~\cite{nihcc} to train a~neural network to extract meaningful features from X-ray images. We trained the~Siamese Network on frontal projections (AP and PA) from each of the~datasets independently and then compared the~obtained results.

Due to the~large imbalance of labels, we decided to select only those that are represented by a~large number of images. From NIHCC~\cite{nihcc}, we excluded the~"Hernia" label with only 227 occurrences, and from PadChest, we decided to use only those that were labeled more than 2,000 times.  It is worth emphasizing that the threshold does not favor any particular labels. However, addressing the challenge of handling a vast number of labels associated with rare occurrences is an independent issue. Finally, 112,010 images and 13 thorax disease categories were used for training on the~NIHCC dataset, and 99,556 images and 27 radiographic findings were selected for training on the~PadChest dataset. The~absence of any label means "no finding" or "normal." The~datasets were divided into training and validation sets in~proportion 4:1.

\begin{figure}[!h]
\centering
\includegraphics[width=0.8\textwidth]{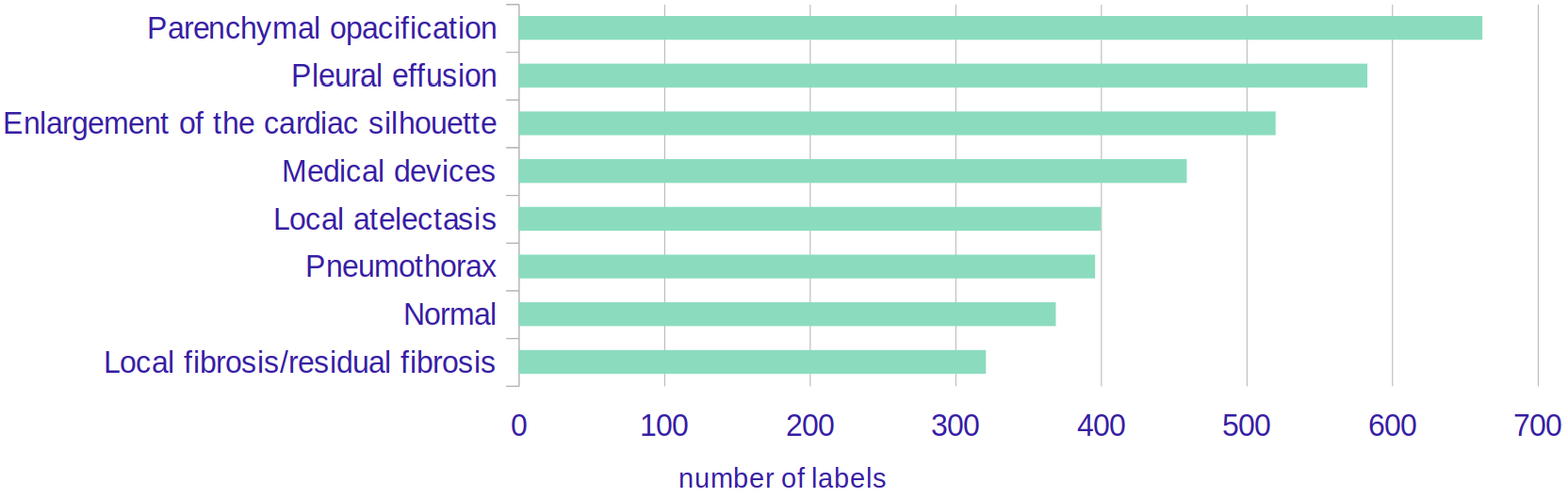}
\caption{The frequency distribution of labels in the~B2000 dataset. The~"Normal" label during training is represented as an absence of lung lesions.}
\label{fig:b2000labels}
\end{figure}

To check the~possibility of transferring meaningful representation by the~Siamese Network, we chose an~internal dataset acquired in~collaboration with the~Medical University of Warsaw called B2000. B2000 contains 2,158 dicom images manually annotated and is suitable for multi-label classification. The distribution of the frequency of occurrence of labels is presented in Fig.~\ref{fig:b2000labels}. This is a~common situation when we have a~small real dataset for a~particular problem and want to get the~most universal representation of it.

\subsection*{Preprocessing of a~grayscale image into 3-channel}

There is a~consensus \cite{checklist} that duplicating channels of a~grayscale image to use a~3-channel pre-trained model unnecessarily increases the~number of~model parameters. Instead, the~remaining two channels can add additional information to the~neural network.

\begin{figure}[!ht]
\centering
\includegraphics[width=\textwidth]{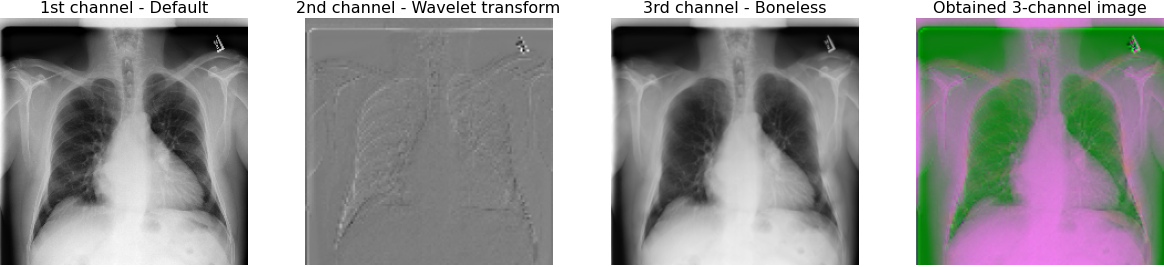}
\caption{Proposed initialization of each image channel and final 3-channel image. The~first channel shows the~original image normalized from 0 to 1. The~second channel is an~image after wavelet transformation. The~third channel is an~image after using the~model to remove bones.}
\label{fig:channels}
\end{figure}

Based on the~work of Matsuyama \cite{Matsuyama2020}, we decided to use wavelet transformation on a~grayscale image. We averaged obtained horizontal, vertical, and diagonal details and set it as a~second channel. As an~input for a~third channel, we used an~image generated by a~deep learning model that returns lungs without ribs \cite{Rajaraman2021}. The~use of this kind of input was suggested by radiologists claiming that, especially on 2D X-ray images, it is important to pay closer attention to areas behind the~ribs to avoid skipping any lesions. It is vital to point out that the~use of additional models in~preprocessing or post-processing can significantly help the~target model perform better. The~example of an~image with channels initialized in the~proposed manner is shown in~Fig.~\ref{fig:channels}.

\begin{figure}[h]
    \centering

        \centering
        \includegraphics[width=0.5\linewidth]{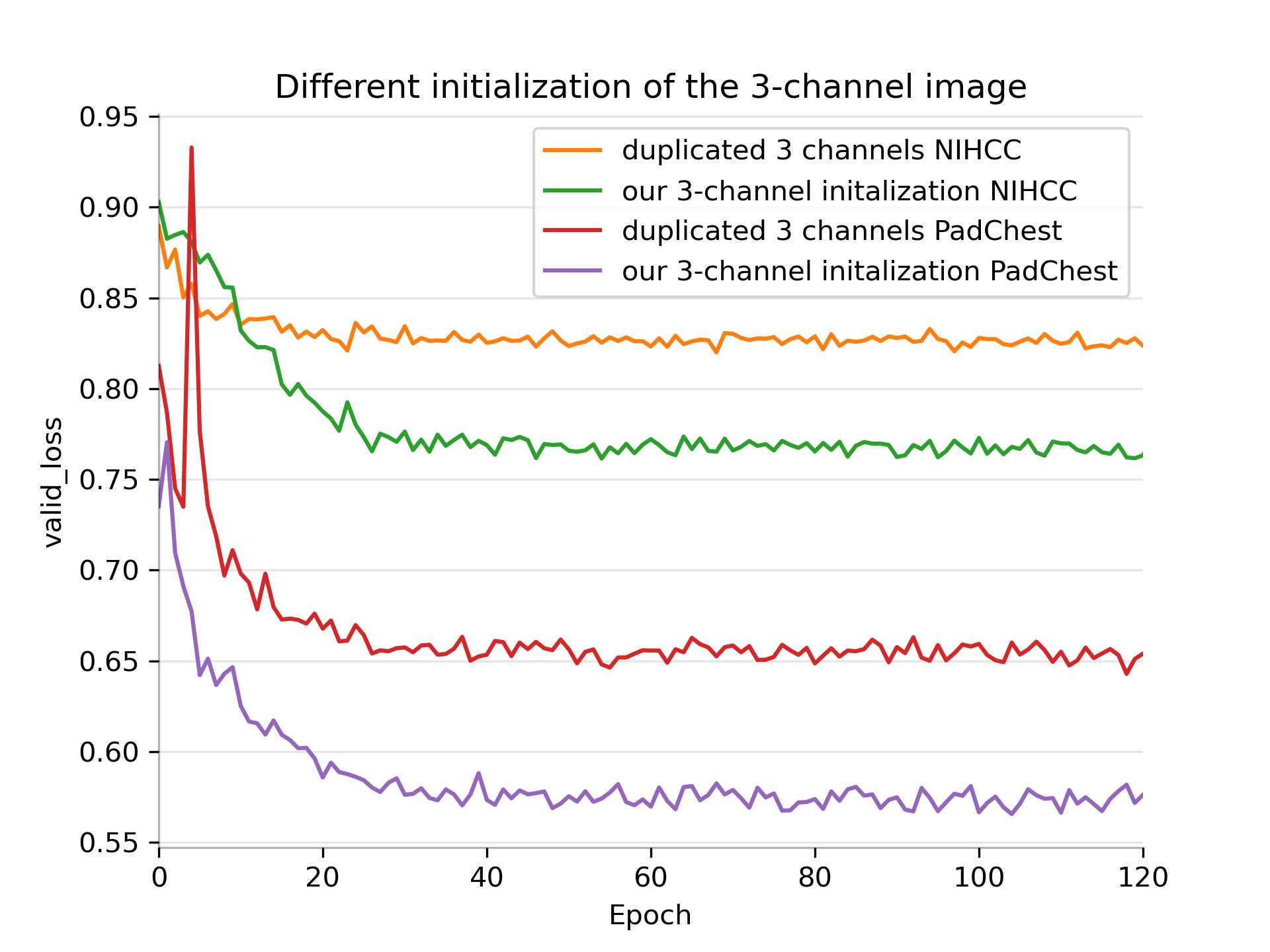}
        \caption{The~validation loss function for the~Siamese neural network with DeiT \cite{deit} as a~backbone with duplicated channels and with channels initialized in~a way shown in~Fig.~\ref{fig:channels}.}
        \label{fig:losschannels}
\end{figure}

During preprocessing, all images were normalized and resized to (380x380) or (384x384) depending on the~model architecture being trained. In our work, we used only the~types of data augmentation that made the~lungs fully visible after the~transformations, for example, randomly shifting (max. 5\%), scaling (max. 5\%), rotating (max. 15\textdegree), adding noise, and distortion. In~Fig.~\ref{fig:losschannels}, the results show the~higher performance obtained by models with our 3-channel initialization procedure.

\begin{figure}[h]
        \centering
        \includegraphics[width=0.8\linewidth]{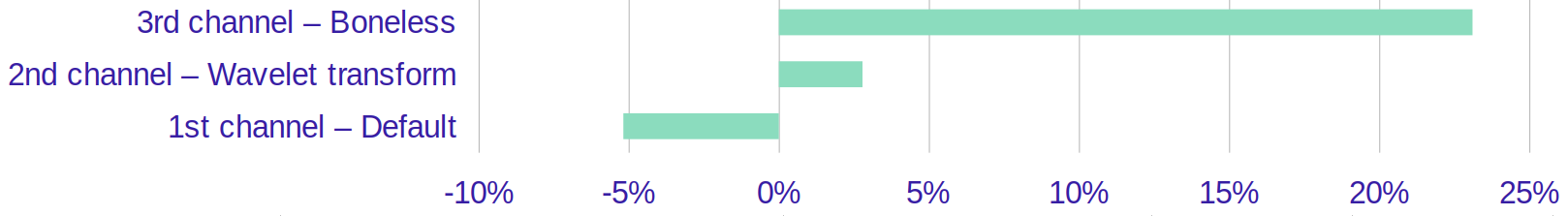}
        \caption{The relevance of a~given channel in a~3-channel image by the~percentage change in accuracy during cross-validation evaluation.}
        \label{fig:channelscontribution}

\end{figure}

To check the~contribution of each of the~proposed channels, we decided to swap one image channel with the~same channel of another randomly selected image. In Fig. \ref{fig:channelscontribution}, the cross-validation results show that in terms of accuracy, the~most important for training is the~channel with images without bones. This~boneless image is the~output of the~deep learning model for bone removal. In comparison with the~proposed 3-channel initialization without any swaps, the~default channel shows a negative impact on the model's performance. Based on this, it can be concluded that the~proposed channel initialization makes the~model learn better.

\subsection*{Algorithm for selecting triplets of images}\label{sec:apnselection}

\begin{algorithm}[h]
    \caption{Algorithm for selecting APN from multi-label dataset}
    \label{alg:algorithm}
    \hspace*{\algorithmicindent} \textbf{Input:} Dataset of images $D$ with labels $L$, a~value showing when to stop looking for more similar or better label $S$ \\
    \begin{algorithmic}[1]
        \For{Anchor image (\textit{A}) in dataset (\textit{D})}
            \For{Positive image (\textit{P}) in randomly sorted dataset (\textit{D}):} \Comment{Find Positive image}

            \State Calculate the set of shared labels (\textit{same}) between \textit{A} and \textit{P}
            \State Calculate the set of differing labels (\textit{diff}) between \textit{A} and \textit{P}
            \State Consider \textit{P} as a potential positive image \textbf{if}:
            \State\;\;\;\; - \textit{A} $\neq$ \textit{P}
            \State\;\;\;\; - \textit{same} is larger than the previously remembered \textit{same}
            \State\;\;\;\; - \textit{diff} is smaller than the previously remembered \textit{diff}
            \State\;\;\;\; - Labels in \textit{A} match labels in \textit{P} \textbf{and} (\textbf{if} \textit{diff} is less than 2 \textbf{or} the negative search has been conducted \textit{S} times) 
                \State\;\;\;\;\;\;\;\; \textbf{then} break
            \EndFor
            
            \State \textbf{do} \Comment{Find Negative image}
            \State\;\;\;\; Randomly select one negative label ($NL_{1}$) for \textit{A}
            \State \textbf{while} $NL_{1}$ is in labels of A \textbf{or} $NL_{1}$ is in labels of previously selected Positive image (\textit{P})
            
            \State \textbf{do}
            \State\;\;\;\; Find random negative image (\textit{N}) with selected label \textit{$NL_{1}$}
            \State\;\;\;\; \textbf{if} searched for Negative image (\textit{N}) \textit{S} times
            
            \State \;\;\;\;\;\;\;\;\textbf{do}
            \State\;\;\;\;\;\;\;\;\;\;\;\;\;  Find another random negative label (\textit{$NL_1$})
            \State \;\;\;\;\;\;\;\; \textbf{while} \textit{NL} is in labels of \textit{A} \textbf{or} \textit{NL} is in labels of previously selected Positive image \textit{P}
            
            \State \textbf{while} $NL$ is in labels of previously selected Negative image \textbf{or} $NL$ is in labels of previously selected Anchor image
        \EndFor
        
    \end{algorithmic}
\end{algorithm}

Creating such Anchor, Positive, and Negative (APN) triplets is a~significant part of Siamese Network training. This is particularly challenging for multi-label data, such as PadChest or NIHCC. Several papers have already attempted this \cite{yang2020, SiameseXML}, but to the~best of our knowledge, there are still no standards for how this should be done.

The~Anchor can be called a~baseline image. Positive and Negative images needed to be very similar and dissimilar to the~Anchor image, respectively. The~selection of pairs, Anchor-Positive and Anchor-Negative, influences further model accuracy. The~considerable size of PadChest and NIHCC made it necessary to create an~effective algorithm to match pairs A-P and A-N. The~proposed algorithm is presented in~Algorithm~\ref{alg:algorithm}. It can be used even for databases that have one-shot learning cases. Another advantage of the~algorithm is the~uniform sampling of examples into A-P and A-N pairs from different labels. It is also worth noting that this~label-balanced algorithm prevents the~same pairs from being selected multiple times thanks to random sorting.

\subsection*{Deep Siamese Network architecture}

\begin{figure}[ht]
\centering
\includegraphics[width=0.6\linewidth]{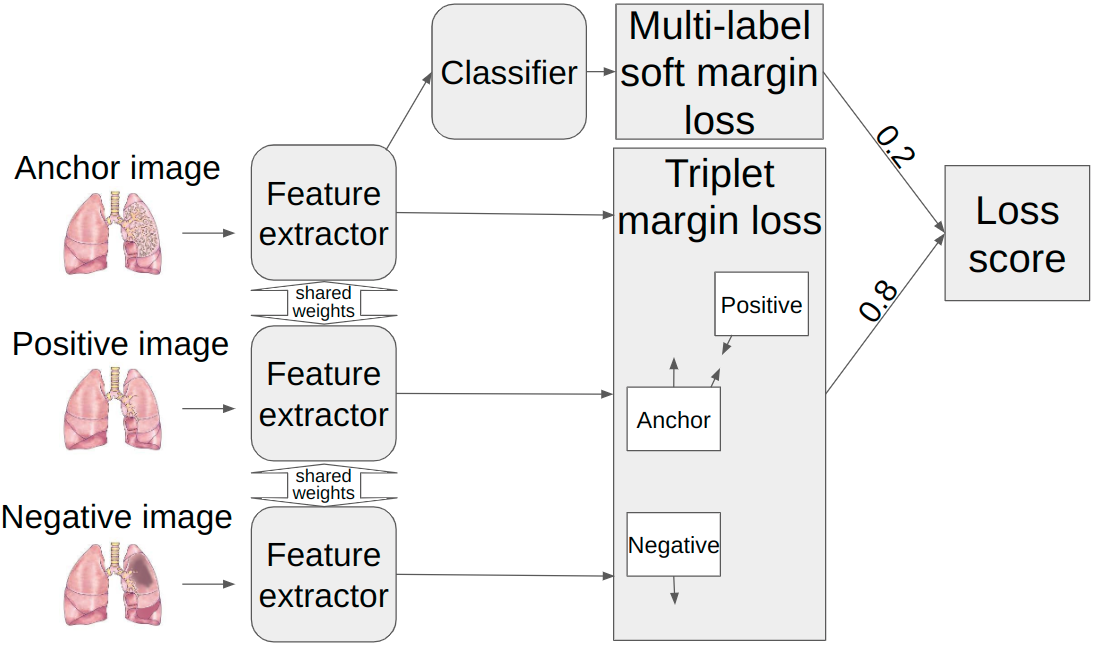}
\caption{Architecture of a~Siamese neural network. Features extractors share the~same weights. The~output of feature extractors is used in two tasks: classification and representation learning. The~losses of these tasks are taken into account to calculate the~final loss score with weights of 0.2 and 0.8.}
\label{figure:siamese}
\end{figure}

Presented in~Fig.~\ref{figure:siamese}, the~architecture of the~Siamese neural network needs three images: Anchor (A), Positive (P), and Negative (N) for each training step. The~APNs are processed through the~Siamese Network, yielding embeddings being created from them. The~embeddings are then given to the~loss function, called triplet loss \cite{Dong_2018_ECCV}, which compares a~baseline image (A) with the~Positive and Negative example, maximizing the~distance between the~Anchor and the~Negative while minimizing the~distance between the~Anchor and the~Positive. A~triplet loss function is defined as
\begin{equation}
\centering
{\displaystyle {\mathcal {L}}\!
\left(A,P,N\right)\!=\!
\operatorname{max}\!
\left(
{\|\operatorname\! {f}\! \left(A\right)\!-\!\operatorname\! {f} \!\left(P\right)\!\|}^{2}\!
-\!
{\|\operatorname\! {f}\! \left(A\right)\!-\!\operatorname\! {f} \!\left(N\right)\!\|}^{2}\!
+\!1\! ,0\right)}.
\label{equation:tripletloss}
\end{equation}

The loss function of the~proposed solution is a~weighted sum of the~loss function derived from the~Siamese Network and the~multi-label classification task of the~Anchor image. Thus, it can be expressed as

\begin{equation} 
\centering
{\displaystyle {\mathcal {L}_{total}}=
0.8\cdot \displaystyle {\mathcal {L}_{siamese}}
\left(A,P,N\right)+
0.2\cdot \displaystyle {\mathcal {L}_{classif}}
\left(A\right).
}
\label{equation:loss}
\end{equation}

Multi-label soft margin loss was selected for the~classification task.

\subsection*{Siamese Network training}

The experiments were conducted to select the~best backbone of the~feature extractor with the~smallest loss value on~the~validation dataset. For each backbone model: MobileNet \cite{mobilenet}, DenseNet \cite{densenet}, EfficientNet \cite{efficientnet}, VIT \cite{vit}, and DeiT \cite{deit}, the~pre-trained weights were uploaded, but the~backbone was not frozen. The~input image size was 380x380x3 or 384x384x3, depending on the~architecture. 

The~Siamese neural network was trained on NVIDIA DGX A100 for 120 epochs with an~early stopping technique and the~batch size equal to 16. Adam optimizer with a~learning rate of 0.0002 and halving the~learning weight every five epochs was applied. The~embedding size was equal to 512.

\begin{figure}[h]
\centering
\includegraphics[width=.49\textwidth]{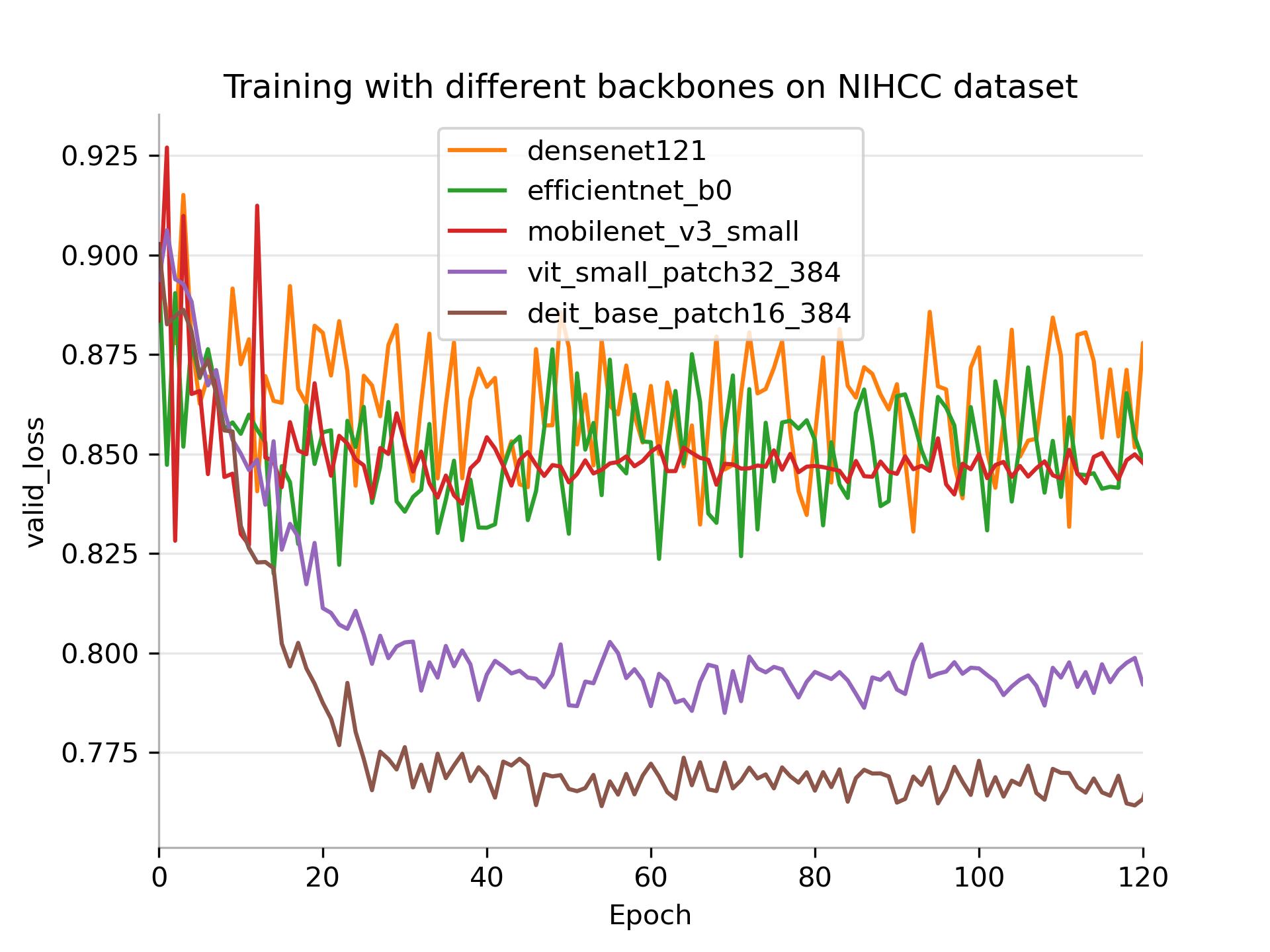}
\includegraphics[width=.49\textwidth]{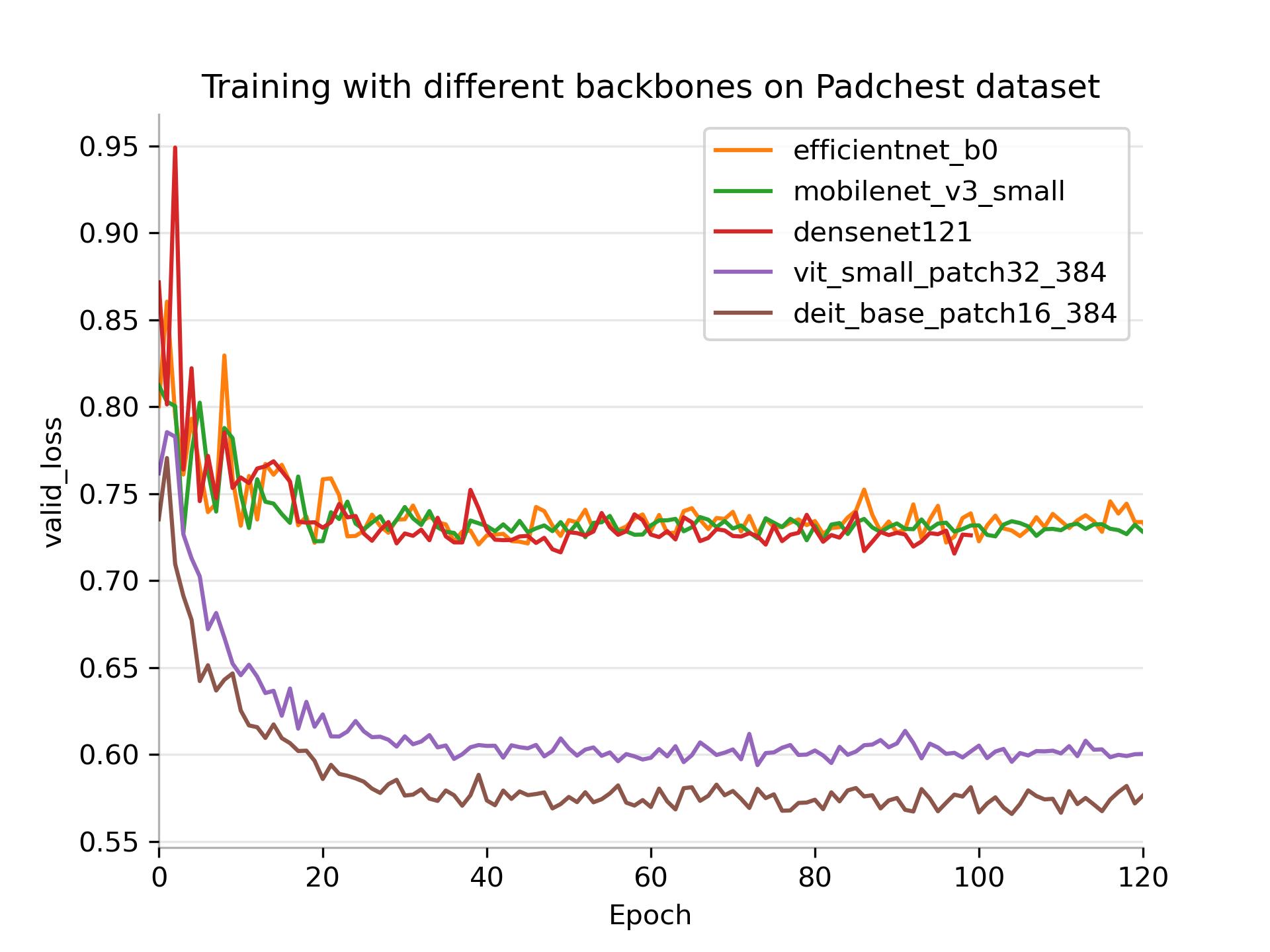}

\caption{The~validation loss function when training the~Siamese neural network on PadChest and NIHCC datasets on the~following backbone networks: MobileNet, DenseNet, EfficientNet, VIT, and DeiT.}  
\label{fig:charts}
\end{figure}

As visible in Fig. \ref{fig:charts}, the~best training results on both datasets (NIHCC and PadChest) were achieved for the~Data-efficient Image Transformer (DeiT). In this model, images are presented as a~sequence of patches of fixed size 16x16. The~total image size is equal to 384x384x3.

\subsection*{Results of the~representation transfer}
Using a~Siamese model trained either on the~PadChest, NIHCC, or B2000 database, we could generate image representation in~the~form of a~vector embedding. Therefore, we converted the~database into embeddings, meaning a~form applicable to train tabular-data machine learning models.

The machine learning classification model XGBoost was evaluated using 5-fold cross-validation. Table~\ref{tab:siamese:XGBoost} compares the~cross-validation metrics obtained on different combinations of training and evaluation databases. It is clearly visible that, despite training the~Siamese Network on B2000 data, evaluation of the~XGBoost model \cite{xgboost} on B2000 data achieved better results when the~Siamese Network was trained on NIHCC data.

\begin{table}[h]
\caption{Results of XGBoost model trained (T) on multi-label data from the~database indicated in~the first row and evaluated (E) on multi-label data from the~database in~the second row.}
\label{tab:siamese:XGBoost}
\centering

\begin{tblr}{colspec={lrrrr}}\hline
{Metric \hspace{7mm} T \\ \hspace{1.75cm} E} & {PadChest \\ PadChest}&  {NIHCC \\ NIHCC}& {PadChest \\ B2000}& {NIHCC \\ B2000}& {B2000 \\ B2000} \\ \hline
Accuracy&0.364&0.386&0.111&\textbf{0.115}&0.108\\
F1 macro&0.024&0.064&\textbf{0.164}&0.159&0.139\\
F1 weighted&0.024&0.317&\textbf{0.167}&0.162&0.144\\
Prec. macro&0.545&0.190&\textbf{0.308}&0.288&0.258\\
Prec. weighted&0.456&0.414&\textbf{0.434}&0.397&0.377\\
Recall macro&0.013&0.058&\textbf{0.185}&0.177&0.161\\
Recall weighted&0.013&0.320&\textbf{0.155}&0.146&0.143\\ \hline
\end{tblr}	
\end{table}

\section*{X-ray representation based on radiomic features} \label{sec:pyradiomics}
\subsection*{Segmentation model training}

\begin{figure}[h]
\centering
\includegraphics[width=0.3\linewidth]{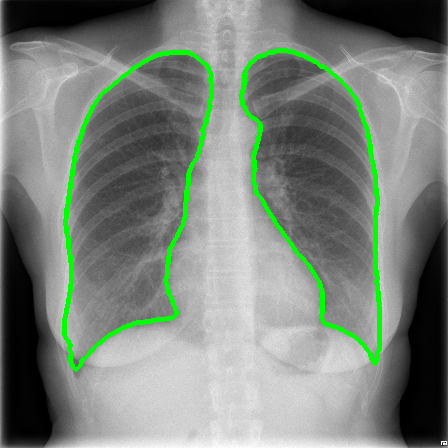}
\caption{Example of segmentation mask generated by CE-Net model \cite{gu2019net} overlaid on X-ray}
\label{figure:mask}
\end{figure}

More accurate radiomic features can be obtained using masks with the~marked relevant areas. For this reason, we decided to obtain such masks by training a~model to segment the~lungs from X-ray images.
As a~dataset for training, we selected MCS~\cite{Jaeger2014} with 704 X-rays and binary masks created for lungs. We allocated 15\% of the~images in~the dataset for validation purposes. 
We chose a~U-net-like architecture, namely CE-Net \cite{gu2019net}, which showed promising results on lung segmentation from 2D CT images, shown in Fig. \ref{figure:mask}. We trained it with images of size (448,448) for 100 epochs, batch size equal to 8, and a~learning rate of 0.0002. The~binary cross-entropy is the~loss function. To enhance generalization accuracy, every ten epochs, the~learning rate was reduced twice.

The model achieved the~best results after 28 epochs of training. Training and validation model metrics are presented in~Table~\ref{tab:metrics:segmentation}.

\begin{table}[h]
\caption{Metrics obtained by segmentation model CE-Net during training and validation.}
\label{tab:metrics:segmentation}
\centering

\begin{tabular}{lrrrrrr}
\hline
& Loss & Accuracy & Precision & Recall & F1 & IoU \\ \hline
Train & 0.042 & 0.983 & 0.971 & 0.962 & 0.967 & 0.936 \\
Valid & 0.050 & 0.981 & 0.964 & 0.963 & 0.963 & 0.929 \\ \hline
\end{tabular}
\end{table}

\subsection*{Radiomic features extraction}

The mask produced by a~well-trained segmentation model can give us much valuable information about the~patient's condition. Due to that fact, we decided to extract features from original images and their masks using the~PyRadiomics library \cite{pyradiomics}. The~library calculates the~features related to, for example, first-order statistics, shape, and texture descriptors. In~our case, it produced 126 features per image. We can use the~obtained features as a~kind of tabular data in~our polyrepresentation learning solution.

\section*{X-ray representation based on self-supervised model}

We leverage the power of DINO \cite{oquab2023dinov2}, a state-of-the-art self-supervised learning method, to obtain embeddings without the need for fine-tuning or weak supervision. DINO is pre-trained on a large, curated dataset using self-supervised learning techniques. The authors claim that this approach allows the model to learn transferable frozen features that closely match the performance of (weakly) supervised alternatives across various benchmark tasks.

\begin{figure}[h]
\centering
  \includegraphics[width=.3\linewidth]{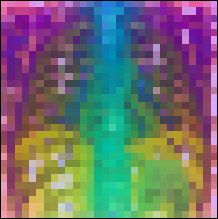} 
\caption{Visualization of DINO features achieved via Principal Component Analysis with a dimensionality reduction to three channels, followed by reshaping image to 30x30 pixels and scaling to a range of [0-255].}
\label{fig:dninoviz}
\end{figure}

An important role in the DINO's pipeline plays preprocessing, consisting of resizing, center cropping, and normalizing. The resulting visual features exhibit remarkable capabilities in understanding object parts and scene geometry across diverse image domains, an~example visualized in Fig. \ref{fig:dninoviz}.

By adopting DINO, we wanted to demonstrate the effectiveness of self-supervised learning in producing competitive visual representations, paving the way for its broader applications in computer vision tasks. 

According to the authors, the larger the DINO model, the better the representation. However, we wanted to limit the embedding size and compare whether models with the same embedding size achieve similar performance metrics. For this reason, we used multidimensional scaling and limited the embedding size to 384, the size of the smallest model. 

Table \ref{tab:dino:comparison} shows that the assessment of which model is the~best depends on the chosen performance metric. Since most of the~metrics had the best results for the largest model ViT-g/14, we chose it.

\begin{table}[h]
    \caption{Comparison of performance of DINO's pre-trained models on multi-label B2000 database.}
    \label{tab:dino:comparison}
    \centering
    \setlength{\tabcolsep}{3pt}
    \begin{tabular}{lrrrrrrrr}
    \hline
    Model & Acc. & F1 macro & F1 weight. & Prec. macro  & Prec. weight. & Recall macro & Recall weight. \\ \hline
    ViT-S/14 distilled & 0.156 & 0.236 & 0.244 & \textbf{0.409} & \textbf{0.553} & 0.241 & 0.220\\
    ViT-B/14 distilled & \textbf{0.161} & 0.236 & 0.231 & 0.373 & 0.467 & 0.253 & 0.208\\ 
    ViT-L/14 distilled & 0.160 & 0.231 & 0.221 & 0.400 & 0.529 & 0.243 & 0.197\\ 
    ViT-g/14 & 0.157  & \textbf{0.254} & \textbf{0.247} & 0.398 & 0.512 & \textbf{0.274} & \textbf{0.223} \\ \hline
    \end{tabular}
\end{table}

\section*{Polyrepresentation learning}

It has been shown repeatedly that ensembling of models works and wins competitions \cite{Miller2020, competit}. Ensemble learning is the~process in~which multiple models are combined to solve a~particular problem. The~aim of using ensemble methods is to improve prediction results so that they are better than any of the~contributing models alone.

In this paper, we introduce the~term \textit{polyrepresentation} for combining models. In~terms of a~polyrepresentation, we call a~representation of the~same object obtained from many different models, perspectives, or data sources, for example, in Fig. \ref{fig:schema}. The~difference between ensembling and polyrepresentation is that the~first generates a~single result, and the~second generates a~representation of data, e.g., in~the~form of vector embeddings.  Moreover, polyrepresentation is created as an early fusion of many data representations, whereas ensemble learning is a late fusion of models' predictions. 

In addition, polyrepresentation further stands out as a modular solution, offering the convenience of flexible customization to meet specific analysis needs. The modular nature of polyrepresentation allows the activation or deactivation of individual representation modules, serving as distinct building blocks that contribute to the overall understanding of the data. Importantly, adjusting the polyrepresentation involves retraining a classic machine learning model rather than a resource-intensive deep learning model. This streamlined process significantly accelerates adaptation and saves computational resources, making polyrepresentation a versatile and efficient choice for dealing with various image-related challenges. In contrast to traditional ensemble learning, which often focuses on generating a single outcome, the~polyrepresentation emphasizes a comprehensive data representation, such as vector embeddings. This innovative approach, exemplified in Fig.~\ref{fig:schema}, encompasses the early fusion of diverse data representations, providing a powerful tool for enhancing understanding and decision-making in various image-based applications.

\subsection*{Training of polyrepresentation}

We created a~polyrepresentation of X-ray lungs. For this reason, we tested on database B2000 how well the~ML model would learn while receiving different features. In~Fig.~\ref{fig:schema}, we showed three sources of training data. The~first one was vector embeddings generated by the~Siamese neural network. The~second one was tabular data. Image databases often contain some metadata, such as age, sex, and medical image-taking parameters. In~our case, we selected only age because that age might be a very good determinant of disease. The~third data source came from the evaluation of a pre-trained self-supervised vision transformer, DINO. The last data source was interpretable features extracted from segmentation masks. The data from different sources were preprocessed by imputing missing values using the~nearest neighbor algorithm and normalizing to [0,1]. 

\begin{table}[h]
\caption{Comparison of classification metrics for various training data configurations. The~best results were bolded.}
\label{tab:polyrep:results}
    \centering
    \begin{tblr}{lrrrrrrr}
      \hline
    & \begin{turn}{90}
    {\tabular{@{}l@{}}Siamese embedding \\Tabular data \\Segmentation features \\ Self-supervised model \endtabular} 
    \end{turn} & \begin{turn}{90}
    {\tabular{@{}l@{}}Siamese embedding \\Segmentation features \\ Self-supervised model \endtabular} 
    \end{turn} & \begin{turn}{90}
    {\tabular{@{}l@{}}only \\Self-supervised model\endtabular} 
    \end{turn} & \begin{turn}{90}
    {\tabular{@{}l@{}}only \\Radiomic features\endtabular} 
    \end{turn} & \begin{turn}{90}
    {\tabular{@{}l@{}} only \\Siamese embedding\endtabular} 
    \end{turn} & \begin{turn}{90}
    {\tabular{@{}l@{}}only \\Tabular data\endtabular} 
    \end{turn} \\   \hline
     Accuracy & 0.153 & \textbf{0.163} & 0.157 & 0.125 & 0.111 & 0.151 \\
     F1 macro & 0.256 & \textbf{0.257} &  0.254 & 0.244 & 0.164 & 0.084 \\
     F1 weighted & 0.256 & 0.256 & 0.247 & \textbf{0.258} & 0.167 & 0.083 \\
     Precision macro & 0.392 & 0.367 & \textbf{0.398} & 0.361 & 0.308 & 0.198 \\
     Precision weighted & 0.494 & 0.410 & \textbf{0.512} & 0.483 & 0.434 & 0.232 \\
     Recall macro & 0.287 & \textbf{0.291} & 0.274 & 0.258 & 0.185 & 0.092 \\
     Recall weighted & 0.239 & \textbf{0.246} & 0.223 & 0.243 & 0.155 & 0.069 \\
       \hline
    \end{tblr}
\end{table}


We trained the~XGBoost model \cite{xgboost} multiple times on different combinations of data sources. Based on Table \ref{tab:polyrep:results}, we can conclude that the~best results were for training on the~self-supervised model, Siamese embeddings, and segmentation features concurrently. It turned out that a~variable such as age did not significantly impact the~classification.

\subsection*{Explanation of polyrepresentation model indicating the~relevance of the~input data}

\begin{figure}[!ht]
\centering
\includegraphics[width=.8\textwidth]{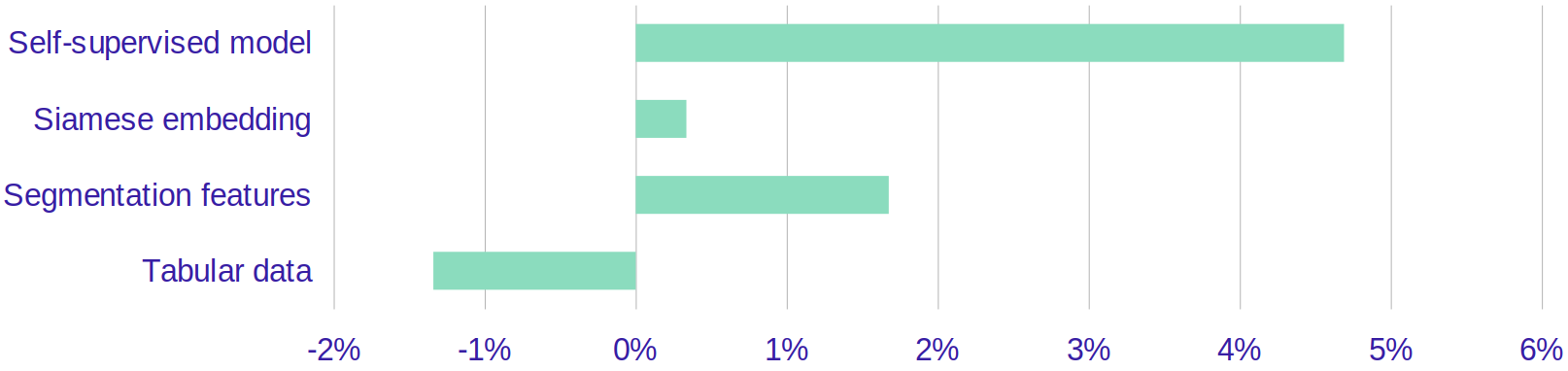}
\caption{The~contribution of four data sources (siamese embeddings, self-supervised visual features, radiomic features, and tabular data) to the~final result, which is denoted by the percentage change in ROC AUC during cross-validation evaluation.}
\label{fig:aspect}
\end{figure}

To confirm our results, we decided to check the~model aspect importance using the~dalex library \cite{dalex}. Fig.~\ref{fig:aspect} confirmed the~importance of Siamese embeddings and segmentation features. However, it also reveals two essential elements; namely, it shows that the~representation obtained by the~self-supervised model is more important in~prediction than segmentation features and siamese embeddings and that tabular data contributes negatively to the~results. Pointing out that prediction based on the patient's age is, however, not a good determinant in our case.

\section*{Conclusions}

In this paper, we showed that transfer learning increased the~predictive performance of the~model verified on a~separate medical dataset, even if that dataset came from a~different medical center (and, in our case, from a~different country with different procedures for performing the~study). We combined three representations into polyrepresentation, which brought better performance results. The~results obtained from the~proposed initialization of three channels confirmed the~worth of adding new information to the~model. The~medical practitioners' suggestion that bones sometimes obscure disease-relevant lesions made boneless channel the~most relevant to the~model.

It is worth noting that polyrepresentation can also be obtained for other types of data, not only for medical ones. The need to choose domain-specific representations (modules) can vary across different domains. Segmentation masks can be replaced with bounding boxes. However, a constraint may be the~lack of available tools similar to the PyRadiomics library for non-medical datasets. The rib removal technique is also specific to lung X-ray images, requiring consultation with domain experts for other medical image types. Furthermore, due to the need to use a model to remove the ribs from the image, the proposed preprocessing phase is more time-consuming than standard image preprocessing.

The advantage of polyrepresentation is that the images can be transformed into vectors in~many supervised, self-supervised, and unsupervised ways. In our case, the~accuracy of representation created using only Siamese Networks is low. Nevertheless, it still makes a positive contribution to the final result. 

We see a need to analyze the~complementarity of a~much larger number of representations combined into a~polyrepresentation. Moreover, due to the multi-faceted view of the data, polyrepresentation can reduce bias; however, further work is needed in this area.

The code for training the~Siamese Network and polyrepresentation is available at \url{https://github.com/Hryniewska/polyrepresentation}.

\section*{Acknowledgements}

This work was financially supported by the Polish National
Center for Research and Development grant number INFOSTRATEG-I/0022/2021-
00, and carried out with the~support of the~Laboratory of Bioinformatics and Computational Genomics and the~High-Performance Computing Center of the~Faculty of Mathematics and Information Science Warsaw University of Technology.

\bibliographystyle{plain} 
\bibliography{references}






\end{document}